\documentclass[letterpaper, 10 pt, conference]{ieeeconf}  

\IEEEoverridecommandlockouts                              
\usepackage[utf8]{inputenc}
\usepackage[T1]{fontenc}
\usepackage{graphicx}
\graphicspath{ {./images/} }

\usepackage{sgame}
\usepackage{amssymb}
\usepackage{amsmath}
\usepackage{xcolor}
\usepackage{float}
\usepackage{comment}
\usepackage{lipsum}
\usepackage{mathbbol}
\usepackage[capitalize]{cleveref}
\usepackage{enumerate}
\usepackage{cite}
\usepackage{mathtools}

\newtheorem{remark}{Remark}
\newtheorem{proposition}{Proposition}
\newtheorem{theorem}{Theorem}

\newtheorem{definition}{Definition}

\newtheorem{lemma}{Lemma}

\newcommand{\mmv}[1]{{\color{magenta}\bf [#1]}}

\DeclareMathOperator{\Equaldef}{\overset{def}{=}}

\title{\LARGE \bf On the role of network structure in learning to coordinate with bounded rationality}

\author{Yifei Zhang and Marcos M. Vasconcelos   
\thanks{Y. Zhang and M. M. Vasconcelos are with the Department of Electrical and Computer Engineering, FAMU-FSU College of Engineering, Florida State University,  Tallahassee, FL 32306, USA. E-mails:
        {\tt   yz23r@fsu.edu, m.vasconcelos@fsu.edu}.}%
}

\begin{document}

\maketitle
\thispagestyle{empty}
\pagestyle{empty}

\begin{abstract}


Many socioeconomic phenomena, such as technology adoption, collaborative problem-solving, and content engagement, involve a collection of agents coordinating to take a common action, aligning their decisions to maximize their individual goals. We consider a model for networked interactions where agents learn to coordinate their binary actions under a strict bound on their rationality. We first prove that our model is a potential game and that the optimal action profile is always to achieve perfect alignment at one of the two possible actions, regardless of the network structure. Using a stochastic learning algorithm known as Log Linear Learning, where agents have the same finite rationality parameter, we show that the probability of agents successfully agreeing on the correct decision is monotonically increasing in the number of network links. Therefore, more connectivity improves the accuracy of collective decision-making, as predicted by the phenomenon known as ``Wisdom of Crowds.'' Finally, we show that for a fixed number of links, a regular network maximizes the probability of success. We conclude that when using a network of irrational agents, promoting more homogeneous connectivity improves the accuracy of collective decision-making.

\end{abstract}

\section{Introduction}

In many networked decision systems, multiple agents collaborate to accomplish a possibly complex task impossible for any individual agent to fulfill. To successfully perform the task without a centralized control station, the agents need to interact with each other to determine whether they will engage with a task presented to them. The interactive process leading to a collective decision to act together can be understood as a \textit{learning to coordinate} algorithm \cite{chamley2004rational}. In an ideal environment, such systems operate assuming the agents are fully rational entities, such as networks of programmable devices such as sensors or robots. However, in many applications where unbounded rationality may not be realistic nor achievable -- biologically inspired swarm robotic systems and social-economic networks are prime examples of such settings. 

Our work is motivated by the role that connectivity plays when agents are learning to coordinate over a network with bounded rationality. Network coordination games \cite{lopez2006contagion} provide a structured framework to study distributed coordination of strategic agents. It has been empirically demonstrated that networks can affect a group’s ability to solve a coordination problem \cite{mccubbins2009connected}.
Using an analytical approach, \cite{Montanari:2010} provided a mathematical foundation to determine what types of networks facilitate the propagation of decisions a network system of homogeneous agents, obtaining a characterization of the rate of convergence as a function of the graph structure. The work in \cite{arieli2020speed} obtained an upper bound on the time for decisions to spread over a network of any size and degree distribution, both directed or undirected. Using a Threshold Model, \cite{rossi2017threshold} provided a recursive equation that approximates the propagation of decisions over large networks of heterogeneous agents. Heterogeneity was also studied in \cite{zhou2023consensus}, where necessary and sufficient conditions where established to ensure perfect coordination within a group of two different classes of agents. Robustness in network coordination is the focus of \cite{arditti2024robust}, in which the agents may be externally influenced by an external field creating a bias towards certain actions. The resilience of strategic networked coordination systems to cybersecurity attacks has been considered \cite{canty2018impact,paarporn2020impact,Paarporn:2021b} and the role of local connectivity in stochastic coordination games \cite[and references therein]{vasconcelos2023coordination} has been studied in \cite{dahleh2016coordination,leister2022social,Wei:2023}. 

The focus of our work is in settings where the agents have bounded rationality, and therefore make imperfect decisions. There is a growing literature of models of bounded rationality, including \textit{prospect theory} and \textit{cognitive hierarchy theory} \cite{sanjab2020game,9992669,9508881,abuzainab2017cognitive}, to name a few. Bounded rationality in the context of \textit{evolutionary game theory} and coordination was studied in \cite{paarporn2023madness}. We specifically consider bounded rationality in the context of \textit{Log Linear Learning} (LLL) \cite{marden2012revisiting}, which implies that actions are chosen with a certain degree of imprecision/randomness, which also propagates over the network preventing coordination to be perfectly achieved, i.e., with probability one. In our previous work \cite{zhang2023rationality}, we have established that in the bounded rationality regime of agents using LLL for coordination over regular networks, there exists a nontrivial trade-off between rationality and connectivity. Therefore, the network structure may affect, for instance, whether the achieving coordination with high probability is possible or not.


In this paper, we extend the analysis in \cite{zhang2023rationality} to irregular graphs. We show that coordination games defined over irregular graphs are exact potential games, and that the probability of successfully learning the optimal Nash equilibrium (the one corresponding to a maximizer of the potential function) is monotone increasing in the number of edges in the graph. In other words, our result shows that lack of rationality can be compensated by increasing the connectivity in the network, akin to a phenomenon popularly known as ``Wisdom of Crowds'' \cite{Becker:2017}. Our second contribution is to establish that, in the  bounded rationality regime, regular networks maximize the probability of successfully learning the optimal equilibrium. Therefore, establishing the result that equal access to connectivity resources improves coordination accuracy.

\section{Problem setup}

We begin by defining the class of binary networked coordination games that will be the focus of this paper. Let $[N]:=\{1,2,\ldots, N\}$ denote the set of agents in the network described by an undirected and connected graph $\mathcal{G}:=([N],\mathcal{E})$.
Two nodes $i,j \in[N]$ are  connected if $(i,j)\in \mathcal{E}$. The set of neighbors of agent $i$ is denoted by $\mathcal{N}_{i} := \{j\in [N] \mid (i,j)\in \mathcal{E}\}$. The number of neighbors of agent $i$ is denoted by $|\mathcal{N}_i|$. We assume there are no self loops, i.e., $(i,i) \notin \mathcal{E},$ $i\in[N]$. 

\subsection{Two-agent binary coordination games}

Let $(i,j)\in\mathcal{E}$, and suppose that $a_i,a_j\in\{0,1\}$ are the actions played by agents $i$ and $j$, respectively. For $\theta \in \mathbb{R}$, the following bimatrix game specifies the payoffs for the pairwise interaction between $i$ and $j$.
\begin{figure}[h!]\hspace*{\fill}%
\centering
\begin{game}{2}{2}[$a_i$][$a_j$]
     & $1$ & $0$             \\
 $1$ & $\Big(1-\frac{\theta}{N},1-\frac{\theta}{N}\Big)$ & $\Big(-\frac{\theta}{N},0\Big)$ \\
 $0$ & $\Big(0,-\frac{\theta}{N}\Big)$ & $\Big(0,0\Big)$  \\
\end{game}\hspace*{\fill}%
\caption[]{A coordination game with parameter $\theta$ between two players.}
\label{bimatrix}
\end{figure}





\vspace{5pt}

\begin{remark}[Payoff interpretation]
The payoff structure of the bimatrix game in \cref{bimatrix} corresponds to a coordination game between two agents. Notice the payoff matrix depends on a parameter $\theta \in \mathbb{R}$, which we refer to as a \textit{task difficulty}. The difficulty is amortized over the total number of agents in the system, $N$. Each agent works on a subtask of difficulty $\theta/N$. We are modeling collaboration, therefore, when agent $i$ decides to take on a subtask, it contributes $1$ unit of effort towards its own subtask and the subtasks of its neighbors.
\end{remark}

\vspace{5pt}

A binary coordination game between two players is characterized by the existence of exactly two pure strategy Nash equilibria: $(0,0)$ and $(1,1)$. The following result establishes the range of values of $\theta$ for which the game in \cref{bimatrix} corresponds to a coordination game.

\vspace{5pt}

\begin{proposition}\label{pure_strategy_equilibria}
    Consider the bimatrix game in \cref{bimatrix}, and $\mathcal{S}$ denote its set of pure-strategy Nash equilibria. 
    The following holds:
\begin{equation}
\mathcal{S} = 
\begin{cases}
\big\{ (0,0) \big\} & \text{if} \ \  \theta > N \\
\big\{ (0,0),(1,1) \big\} & \text{if} \ \ 0 \leq \theta \leq N \\
\big\{ (1,1) \big\} & \text{if} \ \ \theta \leq 0.
\end{cases}
\end{equation}
\end{proposition}

\vspace{5pt}

\begin{proof}
    The proof can be obtained by inspection using the definition of a Nash equilibrium \cite{Hespanha:2017}.
\end{proof}

\vspace{5pt}

Therefore, the range of values of task difficulty that we are interested in analyzing is\textbf{} $\theta \in [0,N]$.


\vspace{5pt}

\subsection{Coordination games over networks}

Extending the two-agent coordination game from \cref{bimatrix} over a network, we obtain the following  \textit{network game} with $N$ agents, where agent $i$ simultaneously plays the same action with all of its neighbors $j\in\mathcal{N}_i$. Let $V_{ij}:\{0,1\}^2\rightarrow \mathbb{R}$ be defined as:
\begin{equation}
V_{ij}(a_i,a_j) := a_i\Big(
a_j-\frac{\theta}{N}\Big).
\end{equation}

In a network coordination game, every agent receives the sum of all the payoffs of the bimatrix games $V_{ij}(a_{i}, a_{j})$  played with each of its neighbors. Therefore for the $i$-th agent, the utility is determined as follows:
\begin{equation}
    U_{i}(a_{i},a_{-i}):= \sum\limits_{j\in \mathcal{N}_{i}} V_{ij}(a_{i},a_{j}).
\end{equation}
Therefore, the payoff of the $i$-th agent in our game is
\begin{equation}\label{eq:payoff_network}
U_{i}(a_{i},a_{-i}) = a_i\Big(\sum_{j\in \mathcal{N}_i}a_j-\frac{|\mathcal{N}_i|}{N}\theta\Big).
\end{equation}
Notice that the payoff in \cref{eq:payoff_network} reflects that more connected nodes face a larger difficulty due to the collaboration with other agents in the network. However, the effort is also aggregated, which leads to a potentially higher utility. 

\section{Potential Network Games}

The class of \textit{potential games} (PG) play an instrumental role in our analysis \cite{monderer1996potential}. PGs enjoy two desirable properties: 1. Every finite PG always admits a Nash equilibrium (NE) in pure strategies; 2. If the agents adhere to LLL on a potential network game, they are is guaranteed to converge to a NE in the limit when their rationality goes to infinity \cite{marden2012revisiting}. 
We will first show that the network game defined by \cref{eq:payoff_network} is an \textit{exact potential game}.
To proceed with our analysis, we must first define this important class of games.

\vspace{5pt}

\begin{definition}[Exact potential games]\label{exact_potential}
Let $\mathcal{A}_i$ denote the action set of the $i$-th agent in a game with payoff functions $U_i(a_i,a_{-i})$, $i\in[N]$. Let $\mathcal{A} = \mathcal{A}_1\times \cdots \times \mathcal{A}_n$. A game is an exact potential game if there is a so-called \textit{potential function} {$\Phi$}: 
 {$\mathcal{A}\rightarrow \mathbb{R}$} such that 
\begin{equation}\label{eq:exact_potential}
    U_{i}(a'_{i},a_{-i})- U_{i}(a''_{i},a_{-i}) = \Phi(a'_{i},a_{-i})- \Phi(a''_{i},a_{-i}),
\end{equation}
for all  $a_{i}',a_{i}''\in \mathcal{A}_{i}$, $a_{-i}\in \mathcal{A}_{-i}$,  $i \in [N]$.
\end{definition}

\vspace{5pt}



We proceed by showing that the game defined by \cref{eq:payoff_network} is
a PG. We start our analysis by restating a classical result by Monderer and Shapley \cite[Theorem 4.5]{monderer1996potential}.


\vspace{5pt}

\begin{lemma}[Monderer and Shapley]\label{thm:monderer}
    Let $\Gamma$ be a game in which the strategy sets are intervals of real numbers. Suppose the payoff functions are twice continuously differentiable. Then $\Gamma$ is a potential game if and only if
\begin{equation}\label{Condition_for_potential}
\frac{\partial^2(U_{i}-U_{j})}{\partial a_i \partial a_j} = 0
\end{equation}
for every $i$,$j$ $\in [N]$.
\end{lemma}

\vspace{5pt}



\par

\vspace{5pt}

\begin{proposition}
    The network game defined by \cref{eq:payoff_network} on any connected undirected graph $\mathcal{G}$ is potential. Let the adjacency matrix of $\mathcal{G}$ be denoted by $\mathbf{A}$. The potential function of the network coordination game is given by
\begin{equation}\label{pot4irregular}
        \Phi(a) = \frac{1}{2} a^{\mathsf{T}}\mathbf{A} a -\frac{\theta}{N} a^{\mathsf{T}}\mathbf{A} \mathbb{1}. 
    \end{equation}
\end{proposition}

\vspace{10pt}

\begin{proof}
Relaxing $a_i \in \mathbb{R} $ for all $i \in [N]$, then the payoff function \cref{eq:exact_potential} is twice differentiable. If the continuous game is potential, then it necessarily implies that the original game is potential, since we can define the potential values on the vertices of the unit lattice in $\mathbb{R}^N$, and assign those as the potential values of the original game. Plugging \cref{eq:exact_potential} into \cref{Condition_for_potential}, we have:
    \begin{equation}
       \frac{\partial^2(U_{i}-U_{j})}{\partial a_i \partial a_j} = \mathbf{1}(i \in \mathcal{N}_j) - \mathbf{1}(j \in \mathcal{N}_i)
    \end{equation}
    where $\mathbf{1}(\mathfrak{S})$ is the indicator function of statement $\mathfrak{S}$.
    Since $\mathcal{G}$ is undirected, then $i\in \mathcal{N}_j \Leftrightarrow  j\in \mathcal{N}_i$. Therefore,     
    \begin{equation}
\mathbf{1}(i \in \mathcal{N}_j) = \mathbf{1}(j \in \mathcal{N}_i),
    \end{equation}
and \cref{thm:monderer} implies that the networked game is potential. 

To obtain a potential function for our network game, define the following function:
\begin{equation}\label{eq:potential_two_agents}
\phi_{ij} (a_i,a_j) := a_ia_j + (1-a_i-a_j)\frac{\theta}{N}. 
\end{equation}
The function $\phi_{ij}$ is a potential function for the two-agent coordination game in \cref{bimatrix}, i.e., 
\begin{equation}
\phi_{ij} (1,a_j) -\phi_{ij} (0,a_j)  = V_{ij}(1,a_j) - V_{ij}(0,a_j),  
\end{equation}
for all $i,j\in [N]$ and $a_j \in \{0,1\}$. Let $\Phi: \{0,1\}^N \rightarrow \mathbb{R}$ as follows:
\begin{equation}\label{potential_function}
    \tilde{\Phi}(a) := \frac{1}{2}\sum\limits_{i\in [N]} \sum\limits_{j\in \mathcal{N}_{i}} \phi_{ij}(a_{i},a_{j}).
\end{equation}
Plugging \cref{eq:potential_two_agents} into \cref{potential_function} and using the fact that the network is undirected, we get:
\begin{equation}\label{eq:potential_candidate}
\tilde{\Phi}(a) = \underbrace{\frac{1}{2}a^\mathsf{T}\mathbf{A} a - \frac{\theta}{N}a^\mathsf{T}\mathbf{A}\mathbb{1}}_{:=\Phi(a)} + \frac{\theta}{2N}\mathbb{1}^\mathsf{T}\mathbf{A}\mathbb{1}. 
\end{equation}
Since the last term in \cref{eq:potential_candidate} is independent of $a$, we may disregard it. Thus, obtaining \cref{pot4irregular}.

\end{proof}
\vspace{5pt}

\section{Network Coordination Games}

Starting from a two-player binary coordination game, we obtained a network game by extending it over a graph, and we showed that it is always potential. 
In this section, we establish that our network game has the \textit{coordination property}, i.e., there are only two NE in pure strategies: $\mathcal{S}=\{\mathbb{0},\mathbb{1}\}$. We proceed by analyzing the potential function in \cref{pot4irregular}.

\vspace{5pt}

\begin{theorem}\label{01irregular}
    For all action profiles $a$ in the action space $\mathcal{A} := \{0,1\}^N$, the potential function $\Phi(a)$ attains its maximum at either $a^{\star} = \mathbb{1}$ or $a^\star = \mathbb{0}$, regardless of the graph structure.
\end{theorem}

\vspace{5pt}

\begin{proof}
    We use the following notation to split an action profile $a$ into three parts: 
    $a = (a_i,a_{\mathcal{N}_i}, a_{-i\backslash \mathcal{N}_i})$, where $a_i \in \{0,1\}$ is the action of $i$-th agent, $a_{\mathcal{N}_i} \in \{0,1\}^{|\mathcal{N}_i|}$ is the action vector of $i$-th agent's neighbors and $a_{-i\backslash \mathcal{N}_i} \in \{0,1\}^{N-1-|\mathcal{N}_i|}$ is the action vector of the remaining agents.
    Consider the following optimization problem:
    \begin{equation}
        \begin{aligned}\label{f_i}
& \underset{a}{\mathrm{maximize}}
& & f_i(a) :=  a_i\Big(\sum_{j\in \mathcal{N}_i} \frac{1}{2} a_j-\frac{|\mathcal{N}_i|}{N}\theta\Big) \\
& \text{subject to}
& & a \in \mathcal{A}.
\end{aligned}
    \end{equation}
Note that $f_i$ is a bilinear function of $a_i$ and $a_{\mathcal{N}_i}$. Therefore, Problem \eqref{f_i} always takes its maximizer in the set $ F_0^i := \{a \in \mathcal{A} \mid a_i = 0 \}$ or the set $ F_1^i :=\{a \in \mathcal{A} \mid a_i = 1, a_{\mathcal{N}_i} = \mathbb{1}_{|\mathcal{N}_i|} \}$. More precisely, we have:
\begin{itemize}
    \item Case 1: If $\theta > \frac{1}{2} N$, then for all $a \in F_0^i$, the objective function $f_i(a)$ attains the maximum value of $0$.
    \item Case 2: If $\theta < \frac{1}{2} N$, then for all $a \in F_1^i$, the objective function $f_i(a)$ attains the maximum value of $(\frac{1}{2} - \frac{\theta}{N})|\mathcal{N}_i|$.
    \item Case 3: If $\theta = \frac{1}{2} N$, then for all $a \in F_0^i \cup F_1^i$, $f_i(a)$ attains the  maximum value of $(\frac{1}{2} - \frac{\theta}{N})|\mathcal{N}_i| = 0$.
\end{itemize}
    For every index $i \in [N]$, the case-by-case analysis above is valid. Notice that the optimality analysis of $f_i$ is independent of the graph structure. Now, we consider the following:
\begin{equation}\label{sumoff_i}
        \begin{aligned}
& \underset{a}{\mathrm{maximize}}
& & \sum\limits_{i=1}^{N}f_i(a) = \frac{1}{2} a^{\mathsf{T}}\mathbf{A} a -\frac{\theta}{N} a^{\mathsf{T}}\mathbf{A} \mathbb{1} \\
& \text{subject to}
& & a \in \mathcal{A}.
\end{aligned}
    \end{equation}
    
     The objective function in \cref{sumoff_i} is precisely $\Phi(a)$. As it is a finite sum of local objective functions, which share the same optimality condition, the optimizer of $\Phi$ should lie in the intersection of the individual optimal sets (if the intersection is nonempty).

     \begin{itemize}
    \item Case 1: If $\theta > \frac{1}{2} N$, then for all $a \in \bigcap_{i=1}^{N} F_0^i$, the objective function $\Phi(a)$ attains the maximum value of $0$. There is only one element in $\bigcap_{i=1}^{N} F_0^i$, which is $\mathbb{0}$. 
    \item Case 2: If $\theta < \frac{1}{2} N$, then for all $a \in \bigcap_{i=1}^{N} F_1^i$, the objective function $\Phi(a)$ attains the maximum value of $(\frac{1}{2} - \frac{\theta}{N})\mathbb{1}^{\mathsf{T}}\mathbf{A} \mathbb{1}$.
    Since the graph is connected, $[N] \subseteq \bigcup_{i=1}^{N} \mathcal{N}_i$, 
    there is only one element in $\bigcap_{i=1}^{N} F_1^i$, which is $\mathbb{1}$.
    \item Case 3: If $\theta = \frac{1}{2} N$, then for all $a  \in \bigcap_{i=1}^{N} F_0^i \cup F_1^i$, the objective function $\Phi(a)$ attains the maximum value of $(\frac{1}{2} - \frac{\theta}{N})\mathbb{1}^{\mathsf{T}}\mathbf{A} \mathbb{1} = 0$. There are two elements in $\bigcap_{i=1}^{N} F_0^i \cup F_1^i$, which are $\mathbb{1}$ and $\mathbb{0}$.
\end{itemize}

In conclusion, $\Phi(a)$ attains its maximum at either $a^\star = \mathbb{1}$ or $a^\star = \mathbb{0}$. Any action profile that contains both 1's and 0's can never be optimal.
\end{proof} 


\vspace{5pt}

The game defined by \cref{eq:payoff_network} is indeed a \textit{network coordination game}, regardless of the graph structure. Moreover, there is a threshold on the task difficulty $\theta^{\mathrm{th}}$, which determines whether $a^\star = \mathbb{0}$ or $a^\star = \mathbb{1}$, given by:
\begin{equation}
\theta^{\mathrm{th}} = \frac{N}{2}.
\end{equation}
Although the graph structure does not play a role in the 
coordination property, it
does affect the properties of the stochastic learning process applied to this game. In the next section, we show how the graph structure influences the probability of all of the agents successfully playing the optimal NE, $a^\star$, using LLL with bounded rationality.

\section{Learning to Coordinate over a Network}

\subsection{Log Linear Learning} We assume that every agent in the network uses a learning algorithm known as \textit{Log Linear Learning} \cite{marden2012revisiting}. LLL is a widely used learning algorithm where each agent updates its action at time $t$ based on its payoff given the actions played by its neighbors at time $t-1$. At each time step {$t>0$}, one agent {$i\in [N]$} is chosen uniformly at random, and allowed to update its current action. All other agents repeat the action taken at the previous time-step. The probability that agent {$i$} chooses action {$a_{i}\in\{0,1\}$} is determined as follows:
\begin{equation}\label{eq:LLL}
   \mathbb{P}\big(A_i(t)=a_{i}\big)= \frac{e^{\beta U_{i}\big(a_{i},a_{-i}(t-1)\big)}}{\sum_{a'_i\in \mathcal{A}_{i}}e^{\beta U_{i}\big(a'_{i},a_{-i}(t-1)\big)}}, \  a_i\in \mathcal{A}_i,
\end{equation}
where the parameter $\beta \geq 0$ captures the \textit{rationality} of the agents. When $\beta \rightarrow 0$, the agents select their actions uniformly at random, whereas, when $\beta \rightarrow \infty$, the agents update their actions using a best response policy to their neighbors actions. If the game is an \textit{exact potential game}, 
then the Markov chain induced by LLL has a unique stationary distribution $\mu: \{0,1\}^N \rightarrow [0,1]$ given by
\begin{equation}
\mu_{\mathcal{G}}(a \mid \beta) :=
\frac{e^{\beta\Phi(a)}}{\sum_{a'\in \{ 0,1\}^N}e^{\beta \Phi(a')}},
\end{equation}
where $\Phi: \{0,1\}^N \rightarrow \mathbb{R}$ is the game's potential function. The stationary distribution gives the probability of the agents playing a certain action profile in the long run, when $t\rightarrow \infty.$ We say that the agents have successfully learned the optimal action profile $a^\star$ if 
\begin{equation}
\lim_{\beta\rightarrow \infty}\mu(a^\star \mid \beta) = 1. 
\end{equation}
However, for $\beta < \infty$, the probability $\mu(a^\star \mid \beta)<1$. In the next section, we investigate if the stationary probability of successful learning can be improved without changing the rationality of the agents, but by augmenting the connectivity of the network instead.



\subsection{Inductive improvement by increasing connectivity}

The simplest metric of network connectivity is the number of edges in the graph used to describe it. In fact, as shown in the next theorem, the steady state probability of LLL learning to play the optimal action profile increases as the number of edges in a graph increases.

\vspace{5pt}

\begin{theorem}\label{inductiveimprove}
Consider an undirected connected graph $\mathcal{G}$, and denote by $\mathcal{G}_s$ an augmented graph of $\mathcal{G}$ such that
\begin{equation}
    \mathcal{G} = ([N],\mathcal{E}) \hspace{5pt}\text{and}\hspace{5pt} \mathcal{G}_s = \big([N],\mathcal{E} \cup (i,j) \big) 
\end{equation}
where agents $i$ and $j$ are not connected in $\mathcal{G}$, i.e., $(i,j) \notin \mathcal{E}$.
For the network coordination game defined by \cref{eq:payoff_network}, the  probability of success using LLL is strictly larger on $\mathcal{G}_s$ than on $\mathcal{G}$ for the same rationality level $\beta$, i.e.,
\begin{equation}
    \mu_{\mathcal{G}_s}(a^\star \mid \beta) > \mu_{\mathcal{G}}(a^\star \mid \beta).
\end{equation}
\end{theorem}

\vspace{5pt}

\begin{proof}
    Let $\mathbf{A}$ and $\mathbf{A}_s$ denote the adjacency matrices of $\mathcal{G}$ and $\mathcal{G}_s$, respectively. Let $\Phi_{s}(a)$ denote the potential function of the action profile $a$, when the underlying graph has adjacency matrix $\mathbf{A}_s$.  We have:
    \begin{equation}
        \mathbf{A}_s - \mathbf{A} = \mathbb{e}_i\mathbb{e}_j^\mathsf{T},
    \end{equation}
    where $\mathbb{e}_i$ and $\mathbb{e}_j$ are the $i$-th and the $j$-th canonical basis vector in $\mathbb{R}^N$.
    We will prove the result for the case $\theta < N/2$, $a^\star = \mathbb{1}$.

    Assuming $\theta<N/2$, \cref{01irregular} guarantees that $a^\star=\mathbb{1}$, for any graph. However, for $a^\star = \mathbb{1}$, the values of the potential function for $\mathcal{G}_s$ and $\mathcal{G}$ satisfy:
\begin{equation}\label{eq:relationship_potential}
        \Phi_s(a^\star) - \Phi(a^\star) = \Big(\frac{1}{2} - \frac{\theta}{N}\Big) \mathbb{1}^\mathsf{T}(e_ie_j^\mathsf{T})\mathbb{1} = \frac{1}{2} - \frac{\theta}{N}.
    \end{equation}

For all $a\in \mathcal{A}$, the following holds:
\begin{equation}\label{eq:pot_relationship}
\Phi_s(a)-\Phi(a) = \Big(\frac{1}{2}-\frac{\theta}{N}\Big)a_ia_j.
\end{equation}
Therefore, for all $a \in \mathcal{A}\backslash \{a\in \mathcal{A} \mid a_i = a_j =1\}$, we have: 
    \begin{equation}
        \Phi_s(a) = \Phi(a). 
    \end{equation}
    Notice that the cardinality of the set $\mathcal{A}\backslash \{a\in \mathcal{A} \mid a_i = a_j =1\}$ is $3\cdot 2^{N-2}$, and therefore, it is never empty. 

Now, consider the following equation:
\begin{equation}
 \begin{aligned}
\mu_\mathcal{G}(a^\star \mid \beta)
&=
\frac{e^{\beta\Phi(a^\star)}}{\sum_{a\in \mathcal{A}}e^{\beta \Phi(a)}}. 
\end{aligned}
\end{equation}
Then, using \cref{eq:relationship_potential}, we have:
    \begin{equation} \label{eq:denominator}
        \begin{aligned}
\mu_\mathcal{G}(a^\star \mid \beta)
&=
\frac{e^{\beta\Phi(a^\star)}e^{\beta(\frac{1}{2}-\frac{\theta}{N})}}{\sum_{a\in \mathcal{A}}e^{\beta \Phi(a)}e^{\beta(\frac{1}{2}-\frac{\theta}{N})}} \\
&=
\frac{e^{\beta\Phi_s(a^\star)}}{\sum_{a\in \mathcal{A}}e^{\beta \big( \Phi(a)+(\frac{1}{2}-\frac{\theta}{N})\big)} }. \\
\end{aligned} 
    \end{equation}

Denote the denominator in \cref{eq:denominator} by $(*)$. The following holds:
\begin{eqnarray}
(*) & = & \sum_{a\in \mathcal{A}}e^{\beta \big( \Phi(a)+(\frac{1}{2}-\frac{\theta}{N})\big)} \\
& = & \sum_{a\in \mathcal{A}\backslash \{a\in \mathcal{A} \mid a_i = a_j =1\}}e^{\beta \big( \Phi(a)+(\frac{1}{2}-\frac{\theta}{N})\big)} \nonumber \\ & &  \qquad + \sum_{a\in \{a \in \mathcal{A}\mid a_i = a_j =1\}}e^{\beta \big( \Phi(a)+(\frac{1}{2}-\frac{\theta}{N})\big)}.
\end{eqnarray}

From \cref{eq:pot_relationship}, we have:
\begin{multline}
(*) = \sum_{a\in \mathcal{A}\backslash \{a\in \mathcal{A} \mid a_i = a_j =1\}}e^{\beta \big( \Phi_s(a)+(\frac{1}{2}-\frac{\theta}{N})\big)} \\  \qquad + \sum_{a\in \{a \in \mathcal{A}\mid a_i = a_j =1\}}e^{\beta \Phi_s(a)}.
\end{multline}
Finally, since $\theta<N/2$, we have:
\begin{equation}
\Phi_s(a) + \Big(\frac{1}{2}-\frac{\theta}{N}\Big) > \Phi_s(a).
\end{equation}
Because $e^{x}$ is strictly increasing and $\mathcal{A}\backslash \{a\in \mathcal{A} \mid a_i = a_j =1\}$ is nonempty, we have:
\begin{equation}
(*) > \sum_{a\in \mathcal{A}}e^{\beta \Phi_s(a)}.
\end{equation}
Therefore,
\begin{equation}
\mu_\mathcal{G}(a^\star \mid \beta) < \frac{e^{\beta\Phi_s(a^\star)}}{\sum_{a\in \mathcal{A}}e^{\beta \Phi_s(a)}}=\mu_{\mathcal{G}_s}(a^\star \mid \beta) .
\end{equation}

 The proof for the other case, $\theta \geq \frac{1}{2}N$, is analogous and thus it is omitted here due to space constraints.




   
\end{proof}
\vspace{5pt}



The main implication of \cref{inductiveimprove} is that we can improve the probability of agents with bounded rationality to successfully learn to coordinate on the optimal NE by increasing the connectivity of the network. 

In \cite[Theorem 3]{zhang2023rationality}, it was established that for every connected $\mathcal{G}$, the probability of successful learning is monotone increasing in the rationality parameter, i.e., 
\begin{equation}
\frac{\partial \mu_\mathcal{G}}{\partial \beta}(a^\star \mid \beta) > 0.
\end{equation}
Since we are interested in the bounded rationality regime, for every graph $\mathcal{G}$ there exists a minimum level of rationality that the agents must have to guarantee a high enough probability of successful learning. We define the following quantity:
\begin{equation}\label{Betamin}
\beta_{\mathcal{G}}^{\min}(\delta) \Equaldef \min \Big\{ \beta \mid \mu_{\mathcal{G}}(a^\star \mid \beta) \geq 1-\delta \Big\},
\end{equation}
where $\delta \in (0,1)$.

From a designer's perspective, we are interested in choosing a network that allows the agents to behave as irrationally as possible as long as they converge to the optimal NE with high probability.

\subsection{Optimality of regular graphs}

In this section, we will answer the following question: Assuming that the agents adhere to updating their strategies according to LLL, for a given $\delta$ and a fixed number of edges $|\mathcal{E}|$, find the graph $\mathcal{G}$ that minimizes $\beta_{\mathcal{G}}^{\min}(\delta)$. We will prove that the class of regular graphs trades-off connectivity for rationality most effectively.



\begin{figure}[t!]
    \centering
\includegraphics[width=1\columnwidth]
{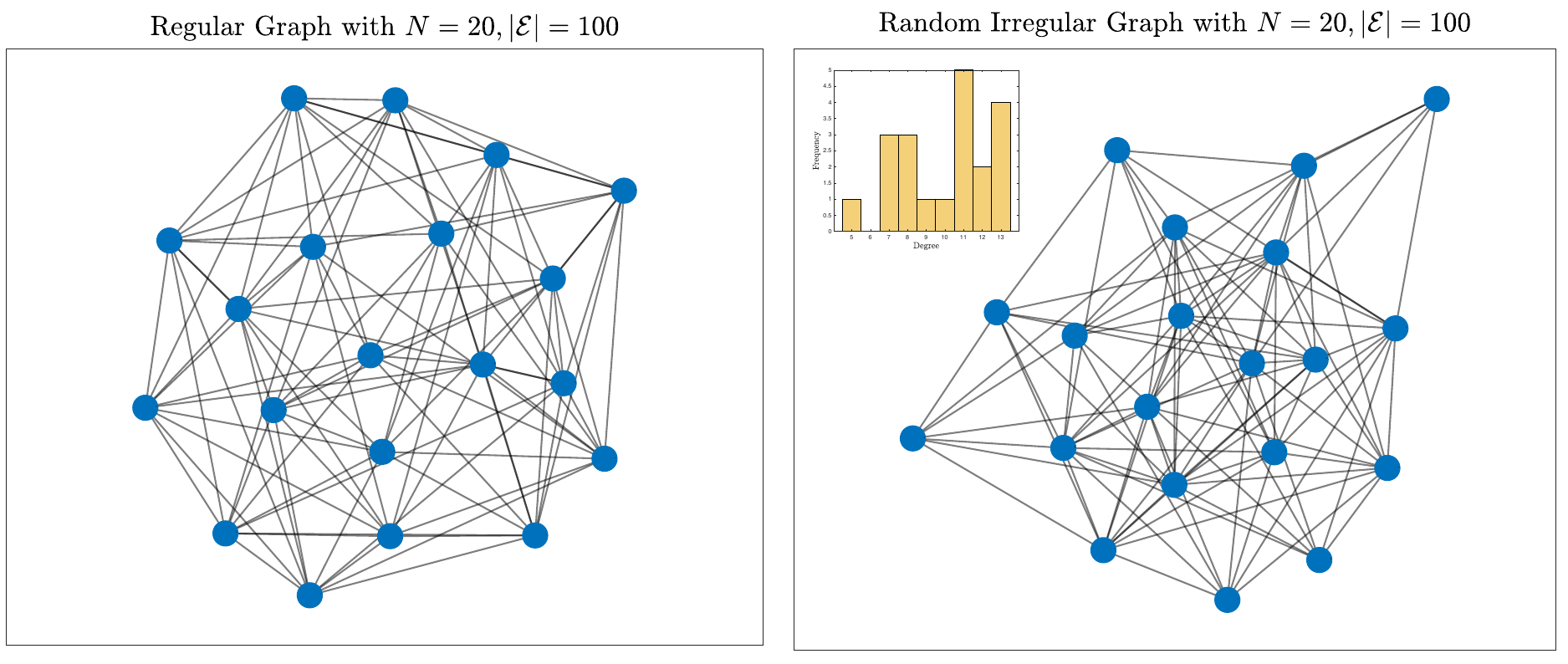}
    \caption{Two graphs with equal number of nodes and edges.}
    \label{fig:graphs}
\end{figure}

A regular graph is one where every node has the same number of neighbors. This is a special class of graphs, which possess interesting robustness properties in a variety of settings \cite{Yaziciouglu:2015}. \Cref{fig:graphs} shows an irregular and a regular graph with the same number of nodes and edges.

Consider two graphs with the same edge connectivity: $\mathcal{G}_R\Equaldef([N],\mathcal{E}_R)$ and $\mathcal{G}_I\Equaldef([N],\mathcal{E}_I)$, where $|\mathcal{E}_R|=|\mathcal{E}_I|$. The subscript $R$ stands for regular and the subscript $I$ stands for irregular. We use other notations in the same manner. For instance, $\Phi_R(a)$ denotes the potential value of an action profile $a$ on a regular graph and $\Phi_I(a)$ denotes the potential value of an action profile $a$ on an irregular graph.

The following lemma states that for our network coordination games, the total potential is invariant on graphs with the same number of nodes and edges.

\vspace{5pt}

\begin{lemma}\label{L1}
Let $\mathcal{G}_R=([N],\mathcal{E}_R)$ and $\mathcal{G}_I=([N],\mathcal{E}_I)$, correspond to a regular and an irregular graph, respectively. Consider the networked coordination game determined by \cref{eq:payoff_network}. If $|\mathcal{E}_R|=|\mathcal{E}_I|$, then
\begin{equation}
    \sum\limits_{a \in \mathcal{A}}\Phi_R(a) = \sum\limits_{a \in \mathcal{A}}\Phi_I(a).
\end{equation}
\end{lemma}

\vspace{5pt}

\begin{proof}
    Consider the map $\rho: \mathcal{A} \to \mathcal{A}$ defined as $\rho(a) \Equaldef \mathbb{1} - a$. Since $\rho(\cdot)$ is a one-to-one correspondence on $\mathcal{A}$, we have:
    \begin{equation}
        \sum\limits_{a \in \mathcal{A}} \Phi(a)+ \sum\limits_{a \in \mathcal{A}}\Phi\big(\rho(a)\big) = 2 \sum\limits_{a \in \mathcal{A}} \Phi(a).
    \end{equation}
   The following identity holds:
    \begin{equation}\label{pairsum}
    \begin{aligned}
         \Phi(a)+\Phi\big(\rho(a)\big) &= \frac{1}{2} a^{\mathsf{T}}\mathbf{A} a -\frac{\theta}{N} a^{\mathsf{T}}\mathbf{A} \mathbb{1} \\
         & \ \ + \frac{1}{2} (\mathbb{1}-a)^{\mathsf{T}}\mathbf{A} (\mathbb{1}-a) -\frac{\theta}{N} (\mathbb{1}-a)^{\mathsf{T}}\mathbf{A} \mathbb{1} \\
         & = \Big(\frac{1}{2}-\frac{\theta}{N}\Big)\mathbb{1}^\mathsf{T} \mathbf{A} \mathbb{1} + a^\mathsf{T} \mathbf{A} (a-\mathbb{1}).
    \end{aligned}       
    \end{equation}
    
    Denote the degree matrix of the associated graph by $\mathbf{D}$. We have $\mathbf{D} = \mathrm{diag} (\mathbf{A}\mathbb{1})$. Observe that for any diagonal matrix $\mathbf{H}$, we have: $a^\mathsf{T} \mathbf{H} (a-\mathbb{1}) = 0$. \Cref{pairsum} becomes:
\begin{eqnarray}\label{Lapquadratic}
\Phi(a)+\Phi\big(\rho(a)\big) & = & 
         \Big(\frac{1}{2}-\frac{\theta}{N}\Big)\mathbb{1}^\mathsf{T} \mathbf{A} \mathbb{1} + a^\mathsf{T} \mathbf{A} (a-\mathbb{1}) \nonumber \\  
            & = &  2\Big(\frac{1}{2}-\frac{\theta}{N}\Big)|\mathcal{E}| + a^\mathsf{T}(\mathbf{A}-\mathbf{D})(a-\mathbb{1})\nonumber \\
           & = &
             2\Big(\frac{1}{2}-\frac{\theta}{N}\Big)|\mathcal{E}|+ a^\mathsf{T}\mathbf{L}(a-\mathbb{1}) \nonumber \\ 
            & = & 2\Big(\frac{1}{2}-\frac{\theta}{N}\Big)|\mathcal{E}|- a^\mathsf{T}\mathbf{L}a, 
\end{eqnarray}
where $\mathbf{L}$ is the graph Laplacian matrix defined by $\mathbf{L} \Equaldef \mathbf{D} - \mathbf{A}$. Note that in the last step to derive Eq. \eqref{Lapquadratic}, we used the fact that $\mathbb{1} \in \mathrm{ker}(\mathbf{L})$ for any undirected unweighted graph.


Equation \eqref{Lapquadratic} implies that for any graph with $|\mathcal{E}|$ edges, we have: 
\begin{equation}\label{Lapsum}
    \sum\limits_{a \in \mathcal{A}} \Phi(a) = 2^N \Big(\frac{1}{2} - \frac{\theta}{N}\Big)|\mathcal{E}| - \frac{1}{2} \sum\limits_{a \in \mathcal{A}} a^\mathsf{T}\mathbf{L}a.
\end{equation}

We proceed to show that 
\begin{equation}\label{sumaLa}
    \sum\limits_{a \in \mathcal{A}} a^\mathsf{T}\mathbf{L}a = 2^{N-1} |\mathcal{E}|.
\end{equation}
To that end, we will use a probabilistic technique. Consider a sequence of $N$ i.i.d. Bernoulli random variables $X_1$, \ldots, $X_N$, where each $X_i$ is distributed according to  $\mathbb{P}(X_i = 1) = \mathbb{P}(X_i = 0) = \frac{1}{2}$. Consider the random vector $X \Equaldef (X_1,\dots,X_N)^\mathsf{T}$. We have:
\begin{equation}\label{defX}
\mathbb{E}[X] = \mathbf{m} =\frac{1}{2}\mathbb{1} \hspace{5pt} \text{and} \hspace{5pt} \mathrm{Var}[X] = \Sigma = \frac{1}{4}\mathbf{I}.
\end{equation}
The random vector $X$ is uniformly distributed on the action space $\mathcal{A}$ with  probability $\frac{1}{2^N}$.
Consider the expectation of the following quadratic form:
\begin{equation}\label{Expectation}
\begin{aligned}
     \mathbf{E}[X^\mathsf{T}\mathbf{L}X] 
     &= \mathrm{trace}\big(\mathbf{L} \cdot \mathbf{E}[XX^\mathsf{T}]\big) \\
     &= \mathrm{trace}\big(\mathbf{L} \cdot (\Sigma + \mathbf{m}\mathbf{m}^\mathsf{T})\big)  \\
     & = \mathrm{trace}(\mathbf{L}\Sigma) + \mathbf{m}^\mathsf{T} \mathbf{L}\mathbf{m}.\\
\end{aligned}
\end{equation}
Since $\mathbf{m} = \frac{1}{2}\mathbb{1} \in \mathrm{ker}(\mathbf{L}) $, we have $\mathbf{m}^\mathsf{T} \mathbf{L}\mathbf{m} = 0$ and 
\begin{equation}
\mathrm{trace}(\mathbf{L}\Sigma) = \frac{1}{4} \mathrm{trace}(\mathbf{L}) = \frac{1}{2} |\mathcal{E}|.
\end{equation}
Thus,
\begin{equation}
     \mathbf{E}[X^\mathsf{T}\mathbf{L}X] = \frac{1}{2} |\mathcal{E}|.
\end{equation}

On the other hand,
\begin{equation}
    \mathbf{E}[X^\mathsf{T}\mathbf{L}X] = \sum\limits_{a \in \mathcal{A}} a^\mathsf{T}\mathbf{L}a \cdot \mathbb{P}(X = a) 
    = \frac{1}{2^N}\sum\limits_{a \in \mathcal{A}} a^\mathsf{T}\mathbf{L}a,
\end{equation}
which implies that:
\begin{equation}\label{sumaLa}
    \sum\limits_{a \in \mathcal{A}} a^\mathsf{T}\mathbf{L}a = 2^{N-1} |\mathcal{E}|.
\end{equation}

Plugging \cref{sumaLa} back into \cref{Lapsum}, we get:
\begin{equation}
    \sum\limits_{a \in \mathcal{A}} \Phi(a) = \Big(\frac{1}{4}-\frac{\theta}{N}\Big)2^{N}|\mathcal{E}|.
\end{equation}
Therefore, the sum of the potential function over the entire action space only depends on the number of edges in the graph. If $|\mathcal{E}_R|=|\mathcal{E}_I|$, the results are the same for $\mathcal{G}_R$ and $\mathcal{G}_I$.
\end{proof}

\vspace{5pt}

Lemma \ref{L1} states that the total potential value only depends on the number of edges. However, the regularity of a graph  affects the distribution of potential values on the vertices of the unit lattice, as stated in the following lemma, which establishes that the potential values are more evenly distributed for regular graphs $\mathcal{G}_R$.

\vspace{5pt}

\begin{lemma}\label{L2}
Let $\mathcal{G}_R=([N],|\mathcal{E}_R|)$ and $\mathcal{G}_I=([N],|\mathcal{E}_I|)$ be (connected) regular and irregular graph with $N$ nodes, respectively. If $|\mathcal{E}_R|=|\mathcal{E}_I| = |\mathcal{E}|$ is such that
\begin{equation}\label{eq:assumption}
 \frac{2|\mathcal{E}|}{N} \in \mathbb{Z}_{\geq 2},
\end{equation}
then
\begin{multline}
\label{Variance}
    \sum\limits_{a \in \mathcal{A}}\Bigg(\Phi_R(a) - \Big(\frac{1}{4}-\frac{\theta}{N}\Big) |\mathcal{E}|\Bigg)^2 \\< \sum\limits_{a \in \mathcal{A}}\Bigg(\Phi_I(a)-\Big(\frac{1}{4}-\frac{\theta}{N}\Big)|\mathcal{E}|\Bigg)^2,
\end{multline}
where $(\frac{1}{4}-\frac{\theta}{N})|\mathcal{E}|$ is the average potential over the entire action space $\mathcal{A}$ for both $\mathcal{G}_R$ and  $\mathcal{G}_I$. 
\end{lemma}

\vspace{5pt}

\begin{proof}
    Consider the uniformly distributed binary random vector $X$ with mean $\mathbf{m}$ and variance $\Sigma$ as in \cref{defX}. The statement in \cref{Variance} is equivalent to:
    \begin{equation}
\mathrm{Var}\big[\Phi_R(X)\big] < \mathrm{Var}\big[\Phi_I(X)\big],
    \end{equation}
    which is also equivalent to:
    \begin{equation}\label{variance}
\mathbf{E}\big[\Phi^2_R(X)\big] < \mathbf{E}\big[\Phi^2_I(X)\big].
    \end{equation}
We proceed by establishing  \cref{variance}. First, we obtain the following expression:
\begin{multline}\label{Var}
    \mathbf{E}\big[\Phi^2(X)\big] 
          = \frac{1}{2} \mathrm{trace}(\mathbf{A}\Sigma\mathbf{A}\Sigma) + \mathbf{m}^\mathsf{T}\mathbf{A}\Sigma\mathbf{A}\mathbf{m} \\
           \hspace{20pt} - \frac{2\theta}{N}\mathbb{1}^\mathsf{T}\mathbf{A}\Sigma\mathbf{A}\mathbf{m} + \frac{\theta^2}{N^2}\mathbb{1}^\mathsf{T}\mathbf{A}\Sigma\mathbf{A}\mathbb{1}. 
    \end{multline}   
    
    Let $\mathbf{d}\Equaldef \mathbf{A}\mathbb{1}$. Note that $\mathbf{d}$ is the degree vector of the associated graph.
After some algebra, we rewrite \cref{Var} as:
\begin{equation}\label{NTN}
    \mathbf{E}\big[\Phi^2(X)\big] = \frac{1}{32}|\mathcal{E}| + \Big(\frac{1}{4} - \frac{\theta}{2N}\Big)^2\|\mathbf{d}\|^2_2.
\end{equation}
Consider the following optimization problem:
\begin{equation}
    \begin{aligned}\label{2-1norm}
& \underset{\mathbf{d} \in \mathbb{Z}^N}{\mathrm{minimize}}
& &   \|\mathbf{d}\|_2 \\
& \text{subject to}
& &\mathbf{d} \succeq \mathbb{1} \\
& & &\mathbb{1}^\mathsf{T}\mathbf{d} = 2 |\mathcal{E}|.
\end{aligned}
\end{equation}

To solve this problem, we rewrite the equality constraint as follows:
\begin{equation}
\mathbb{1}^\mathsf{T}\mathbf{d} = \| \mathbb{1}\|_2\|\mathbf{d}\|_2\cos(\tau)= 2 |\mathcal{E}|, 
\end{equation}
where $\tau$ denotes the angle between the vectors $\mathbf{d}$ and $\mathbf{1}$. Therefore, the minimum norm vector $\mathbf{d}^\star$ is obtained by making $\cos(\tau) = 1$, which implies that 
\begin{equation}
\mathbf{d}^\star=\alpha^\star \mathbb{1}, \ \ \alpha^\star>0.
\end{equation}
Therefore,
\begin{equation}
\alpha^\star = \frac{2|\mathcal{E}|}{N}.
\end{equation}
Our assumption in \cref{eq:assumption}, guarantees the feasibility of $\mathbf{d}^\star=\alpha^\star\mathbf{1},$ which implies that the graph that minimizes \cref{Var} is regular.
\end{proof}

\vspace{5pt}

Equipped with Lemmas \ref{L1} and \ref{L2}, we are ready to present our final Theorem, which qualifies a regular graph as the best candidate for LLL in other to achieve the smallest $\beta^{\min}_{\mathcal{G}}(\delta)$.

\vspace{5pt}

\begin{theorem}\label{RvsI}
    Let $\mathcal{G}_R$ and $\mathcal{G}_I$ be a regular graph and an irregular graph with the same number of vertices and edges, respectively. Suppose LLL is performed on both networks with the same rationality level $\beta$. The corresponding steady-state success probabilities satisfy:
    \begin{equation}
        \mu_R(a^\star \mid \beta) > \mu_I(a^\star \mid \beta)
    \end{equation}
\end{theorem}

\vspace{5pt}

\begin{proof}
We prove the result for $a^\star = \mathbb{1}$. The other case, $a^\star = \mathbb{0}$, is analogous, and it is omitted for brevity. Consider the following quantities:
    \begin{equation}\label{ProbR}
        \mu_R(a^\star \mid \beta)=
\frac{e^{\beta\Phi_R(a^\star)}}{\sum_{a \in \mathcal{A}}e^{\beta \Phi_R(a)}}
    \end{equation}
    and
    \begin{equation}\label{ProbI}
        \mu_I(a^\star \mid \beta)=
\frac{e^{\beta\Phi_I(a^\star)}}{\sum_{a \in \mathcal{A}}e^{\beta \Phi_I(a)}}.
    \end{equation}
    Comparing the numerators of \cref{ProbR} and \cref{ProbI}, we find that they are equal, since:
    \begin{equation}
    \Phi_R(\mathbb{1}) = 2 \Big(\frac{1}{2} - \frac{\theta}{N}\Big) |\mathcal{E}| = \Phi_I(\mathbb{1}).
    \end{equation}
    
    Before we proceed to evaluate the denominators, consider the following optimization problem:
    \begin{equation}
         \begin{aligned}\label{sumofexp}
& \underset{x_1,\dots,x_M}{\mathrm{minimize}}
& &   \sum\limits_{i=1}^{M} e^{\beta x_i} \\
& \text{subject to}
& &  \sum\limits_{i=1}^{M} x_i = c, \\
\end{aligned}
    \end{equation}
    where $c$ and $\beta$ are constants.
  The unique minimizer of Problem \eqref{sumofexp} is 
\begin{equation}
x^\star_1 = \dots = x^\star_M = \frac{c}{M}.
\end{equation}

 Using Majorization Theory \cite{marshall1979inequalities}, the following result can be derived: Consider the set $\mathcal{C}\Equaldef \{ x\in \mathbb{R}^M \mid x_1 + \cdots + x_M =c \}$. Let $\mathbf{x},\mathbf{y}\in \mathcal{C}$. If
    \begin{equation}
        \sum\limits_{i=1}^{M} \Big(x_i-\frac{c}{M}\Big)^2 \leq  \sum\limits_{i=1}^{M} \Big(y_i-\frac{c}{M}\Big)^2 
    \end{equation}
    then 
    \begin{equation}
        \sum\limits_{i=1}^{M}  e^{\beta x_i} \leq \sum\limits_{i=1}^{M}  e^{\beta y_i}. 
    \end{equation}
    
    
    
    \par
For $\mathcal{G}_R$ and $\mathcal{G}_I$ with the same number of edges, from Lemma \ref{L1}, we have: 
\begin{equation}
    \sum\limits_{a \in \mathcal{A}}\Phi_R(a) = \sum\limits_{a \in \mathcal{A}}\Phi_I(a),
\end{equation}
and, from Lemma \ref{L2}, we have:
\begin{multline}
    \sum\limits_{a \in \mathcal{A}}\bigg(\Phi_R(a) - \Big(\frac{1}{4}-\frac{\theta}{N}\Big) |\mathcal{E}|\bigg)^2 \\ < \sum\limits_{a \in \mathcal{A}}\bigg(\Phi_I(a)-\Big(\frac{1}{4}-\frac{\theta}{N}\Big)|\mathcal{E}|\bigg)^2.
\end{multline}
Let $M = 2^N$, $c = (\frac{1}{4}-\frac{\theta}{N})2^{N}|\mathcal{E}|$ and $x_i$'s are the potential values at the action profiles $a\in\mathcal{A}$. Plugging these back into Problem \eqref{sumofexp}, it is immediate to see that:
\begin{equation}
    \sum_{a \in \mathcal{A}}e^{\beta \Phi_R(a)} < \sum_{a \in \mathcal{A}}e^{\beta \Phi_I(a)}
\end{equation}
since on the action space $\mathcal{A}$, a regular graph has the most evenly distributed potential values compared to any irregular graph. Finally, notice that $\mu_R(a^\star \mid \beta)$ and $\mu_I(a^\star \mid \beta)$ have the same numerator, while the latter has a strictly larger denominator. Therefore,
\begin{equation}
    \mu_R(a^\star \mid \beta) > \mu_I(a^\star \mid \beta).
\end{equation}
\end{proof}

\vspace{5pt}

\begin{remark} Theorem \ref{RvsI} implies that any irregular graph with the same number of edges as a regular graph yields a lower probability of learning success when the agents have the same bounded rationality level. In practical settings such as the design of robotic, sensor or satellite networks, the cost of operating the system is likely to be increasing in the number of edges in the graph. Thus, if learning is to be performed, we should always choose to employ a regular network structure if the parameters allow it. In the case that $2|\mathcal{E}|/N$ is not an integer, we should design the network according to a ``near-regular'' graph, such that most of the nodes have the same degree.


\end{remark}

\section{Numerical Results}

\begin{figure}[t!]
    \centering
\includegraphics[width=1\columnwidth]
{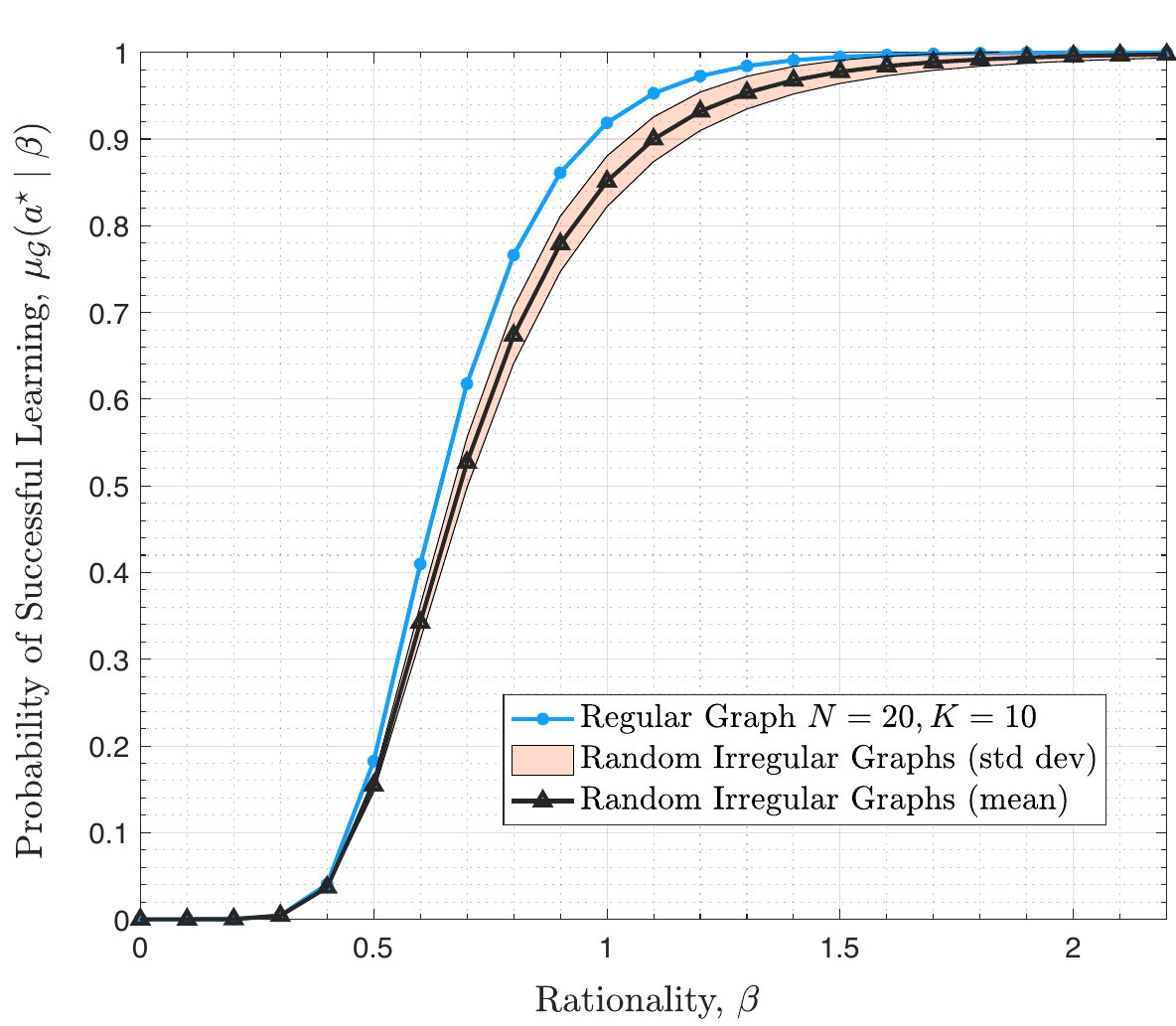}
    \caption{Most papers should have a block diagram that summarizes the problem we are addressing.}
    \label{fig:learning}
\end{figure}

To provide a visual representation of our results, consider a network coordination game with $N=20$ agents, over connected graphs with $|\mathcal{E}|=100$ edges, and parameter $\theta=9$. Since $\theta < \theta^{\mathrm{th}} = N/2=10$, the optimal strategy according to \cref{{01irregular}} is $a^\star=\mathbb{1}$. The probability of successfully learning $a^\star$ as a function of the rationality parameter $\beta$ for different graphs is shown in \cref{fig:learning}. Our base line for comparison in \cref{fig:learning} is the regular graph, whose probability of learning success is strictly larger that any of the other 100 randomly generated graphs obtained for this simulation. The average performance and the standard deviation of these randomly generated graphs is shown in the shaded region strictly below the curve for the regular graph. This is an illustration of \cref{RvsI}, which holds for any number of nodes and edges, provided that there exists a regular graph with such parameters.  

\section{Conclusions and future work}

Network coordination games have been used to model many interesting applications in economics and engineering. A large number of works focus on the convergence analysis of learning in settings with perfectly rational agents. Instead, we studied the case when the agents have bounded rationality, and the effect that connectivity in the network structure has on the probability that the system will converge to the optimal Nash equilibrium solution. We have shown that connectivity can be used to compensate for the lack of rationality, in the sense that adding communication links increase the probability of convergence. Then, we proved that among all irregular networks with the same number of nodes and edges, regular networks (when one exists), have a strictly larger probability of success that irregular networks. The design implication is that, when the system designer is interested in enhancing learning success, it should enable equal access to connectivity by every agent.

There are many interesting research directions stemming from this work. The first is to introduce randomness in this game, by considering a Bayesian setting with random task difficulty $\theta$. A second would be to consider a heterogeneous setting where the agents engage in LLL (or an equivalent algorithm) with different levels of rationality. Finally, we would like to consider the coordination problem with multiple tasks where agents have different affinities over tasks, and study the effect of network structure in learning in this case.

\bibliography{./reference/ref}
\bibliographystyle{ieeetr}

\end{document}